# Algorithm as Defining Dynamic Systems

Keehang Kwon
Department of Computer Engineering
Dong-A University
Busan, Republic of Korea

Hong Pyo Ha
Department of Computer Engineering
Dong-A University
Busan, Republic of Korea

*Abstract*—This paper proposes a new view to algorithms: Algorithms as defining dynamic systems. This view extends the traditional, deterministic view that an algorithm is a step-by-step procedure with nondeterminism. As a dynamic system can be designed by a set of its defining laws, it is also desirable to design an algorithm by a (possibly nondeterministic) set of defining laws.

This observation requires some changes to algorithm development. We propose a two-step approach: the first step is to design an algorithm via a set of defining laws of dynamic system. The second step is to translate these laws (written in a natural language) into a formal language such as linear logic.
key words: dynamic, systems, algorithm, nondeterminisim, linear logic.

I. INTRODUCTION

Designing an algorithm is central to the development of software. For this reason, many algorithms have been developed. However, no guidelines for designing an algorithm have been provide so far: this deficiency is mainly due to the lacking this understanding, algorithm are being designed in an ad-hoc fashion. As a consequence, designing algorithms has been quite cumbersome and error-prone.

What is software/algorithm? Computer science is still looking for an answer to this question. One attempt is based on the view that software is a function and an algorithm is a sequence of instruction for implementing the function. This view has been popular and adopted in many algorithm textbook[6]. Despite some popularity, this view of sequential algorithms stands for a deterministic computation and lack devices for handling nondeterminism. Lacking such devices as nondeterministic transitions, dealing with nondeterminism in this view is really difficult and relies on extra devices such as stacks(for DFS) and queues (for BFS). Those extra devices greatly reduce the readability and modifiability of the algorithm.

This paper proposes another view of software/algorithms, i.e., software as (possibly nondeterministic) dynamic systems and algorithms as defining dynamic systems.
This paper also considers its effects on the algorithms development process. To be precise, we consider here algorithm design to be the process of finding a set of defining laws of dynamic system.

An attractive feature of this view is that it enhances the readability and modifiability of the algorithm for nondeterministic problems. The remainder of this paper is structured as follows. We discuss a new way of describing algorithms in the next section. In Section 3, we present some examples. Section 4 concludes the paper.

II. ALGORITHMS AS DEFINING DYNAMIC SYSTEMS

Our interest is in a process for developing algorithms based on the observation describe in the previous section. The traditional, sequential algorithm process models provide a useful structure for such a process, but some changes are needed. The first problem arises from the machine-dependent, deterministic view for algorithms. A standard definition is that an algorithm is a sequence of instructions. This definition requires algorithms to be deterministic. However, it is easily observed that this deterministic view makes an algorithm to be (sequential) machine-dependent and extra complicated. In algorithm design, nondeterministic algorithms are desirable quite often. This natural when there are multiple ways to get there and we simply don't know in advance which of them is chosen. Such examples include graph algorithms, backtracking algorithms, and AI planning problems.

In ensuring that algorithms are described as simple and machine-independent as possible, it is desirable to express an algorithm via a set of governing laws- in natural language – in the form of initial resources and transition rules. In fact, the above approach to defining algorithms has been used for centries in other fields such as physic and mechanics. The second problem arises from the specification languages to translate these laws. In choosing a language, there is an aspect that requires a special attention. First, we observe that translating the laws into a sequential pseudo code makes the resulting description much bigger, leading to extra complications. An acceptable language should not expand the resulting description too much, rather support a reasonable translation of the laws. An ideal language would support an optimal translation of the laws. Unfortunately, it is a never-ending task to develop this ideal language, as there are too many dynamic systems with too many different features: autonomous systems, open systems with interactions, stochastic systems, etc. We argue that a reasonable, high-level translation of the laws can be achieved via linear logic[3]. An attractive feature of linear logic over other formalisms such as nondeterministic Turing machines, recursive functions, sequential pseudo code, etc, is that it can optimally encode a number of essential characteristics of dynamic system: nondeterminism, updates (also called state change), etc. Hence, the main advantage of linear logic over other formalisms is the minimum (linear) size of the encoding of governing laws of most dynamic systems. The basic operator in linear logic is the linear implication of the form a–b. This expression means that the resource A can be transformed to another resource B. The expression A⊗B means two resources A and B. The expression !A means the resources A is reusable. We point the reader to [3] to find out more about the whole calculus of linear logic.

We sum up our observation in the following equation:

software        = dynamic system.
algorithm design   = a set of defining laws.
algorithm writing  = translation of defining laws.
                     into linear logic.

III. EXAMPLES

The view of "software-as-dynamic-systems" makes algorithms simpler and versatile compared to traditional approach. As an example, we present the factorial algorithm to help understand this notion. The factorial algorithm can be seen as a dynamic system consisting of two laws described below in English:

(1) Initial resource (0, 1).

(2) Transition: $(X, Y)$ can be replaced by $(X+1, XY+Y)$.

This algorithm discards the old resource to produce the new resource and is, therefore, more efficient in term of space usage than its Prolog counterpart. It is shown below that the above laws can be translated into linear logic formulas of the same size. A state is described by a collection of resources. A resource a is represented by a linear logic formula of the form $d(a)$ is represented by a linear logic formula of the form(d) where a is a resource under a directory d. For example, $fact(0,1)$ represents the fact that there exist a resource (0,1) under the directory *fact*. The following is a linear logic translation of the above algorithm, where the reusable action is preceded with !.

   $fact(0,1)$.

   ! ($fact(X,Y)$ –$fact(X+1,XY+Y)$).

A final state is typically given by a user in the form of a query. Computation tries to solve the query. As an example, solving the query $fact(5,X)$ would result in the initial resource $fact(0,1)$ being transformed to $fact(1,1)$, then to $fact(2,2)$, and so on. It will finally produce the desired result $fact(5,120)$ using the second law five times. We now consider the problem of finding the maximum value of the n elements. Suppose they are 5, 10, 9, 2. The standard algorithm creates a new directory max where it keeps track of the maximum value of the elements .An alternative, more dynamic algorithm is shown below:

(1) Initial resources: 4 elements consisting of 5,10,9,2.

(2) Transitions: pick two elements *p* and *q*, and discard the smaller one.

This algorithm produces the desired output by repeatedly *discarding* the smaller input resources. The following is a linear logic translation of the above algorithm.

   $i(5) \otimes i(10) \otimes i(9) \otimes i(2)$.
   $!((i(X) \otimes i(Y) \otimes <(X,Y)) - i(Y))$.
   $!((i(X) \otimes i(Y) \otimes \geq (X,Y)) - i(X))$.

Note that the fact that 3 is an item is encoded as the proposition $i(3)$, i.e., there is a file whose name is 3 under directory i. We assume that, in dealing with $<(X,Y)$, each file $(X,Y)$ such that X is smaller than Y will be created dynamically under the directory $<$. A final state is a state where there is only one element remaining. Hence, solving the query $i(X)$ will produce $i(10)$ – after deleting $i(5)$, $i(9)$ and $i(2)$ – using the second law three times. It is observed that this kind of algorithm is not easily translated into a sequential pseudo code, as the pseudo code has no construct for discarding the input resources. A good motivation for introducing the nondeterminism might be graph algorithms. An example of nondeterministic problems is provided by the following which computes connectivity over an infinite, directed graph. Now we try to determine whether the string *miuiuiu* can be produced from mi with the following four rules:

(a) If you possess a string of the form *Xi*, you can replace it by *Xiu*.

(b) Suppose you have *mX*. Then you can replace it by *mXX*.

(c) A string of the form *XiiY* can be replaced by *XuY*.

(d) A string of the form *XuuY* can be replaced by *XY*.

This problem requires both nondeterminism (There are multiple paths from a node) and updates (An old node is replaced by a new one). For example, the string *mi* can become either *miu* or *mii*. An algorithm for this problem based on functions would be awkward, as functions are too weak, i.e., they support neither nondeterminism nor updates. On the other hand, an algorithm for this problem can be easily formulated as a nondeterministic dynamic system with the following five laws:

(1) Initial resource: *mi*.

(2) Transition: if *Xi*, you can replace it by *Xiu*.

(3) Transition: if *mX*, you can replace it by *mXX*.

(4) Transition: if *XiiY*, you can replace it by *XY*.

(5) Transition: if *XuuY*, you can replace it by *XY*.

Note that this algorithm does not concern whether it will use DFS or BFS when it explores the graph. The following is a linear logic translation of the above algorithm.

s(*mi*).

!∀X(s(*Xi*)⊸s(*Xiu*)).

!∀X(s(*mX*)⊸s(*XX*)).

!∀X(s(*XiiiY*)⊸s(*XuY*)).

!∀X(s(*XuuY*)⊸s(*XY*)).

Now solving the query s(*miuiuiu*) would decide whether *miuiuiu* can be produced in the above puzzle.

Another example of nondeterministic problem is provided by the following menu at a fast-food restaurant. Now we try to determine what can be obtained for four dollars.

(a) three dollars for a hamburger set consisting of a hamburger and a coke,

(b) four dollars for a fish burger set consisting of a fish-burger and a coke,

(c) three dollars for a hamburger, four dollars for a fish burger, one dollar for a coke (with unlimited refills), and one dollar for a fry.

The following is a linear logic translation of the above algorithm.

*p(4)*.

!∀X(*p(X)* ⊗ ≥(*X*,3)⊸ *p(h)* ⊗ *p(c)* ⊗ *p(X-3)*).

!∀X(*p(X)* ⊗ ≥(*X*,4)⊸ *p(fi)* ⊗ *p(c)* ⊗ *p(X-4)*).

!∀X(*p(X)* ⊗ ≥(*X*,3)⊸ *p(h)* ⊗ *p(X-3)*).

!∀X(*p(X)* ⊗ ≥(*X*,4)⊸ *p(fi)* ⊗ *p(X-4)*).

!∀X(*p(X)* ⊗ ≥(*X*,1)⊸ *p(c)* ⊗ !*p(c)*⊗*p(X-1)*).

!∀X(*p(X)* ⊗ ≥(*X*,1)⊸ *p(f)* ⊗ *p(X-1)*).

The proposition *p*(4) represents that a person has four dollars. Now solving the query *p(h)* ⊗*p*(c) ⊗*p*(f) would succeed as we can obtain a hamburger and a coke for three dollars, and a fry for a dollar. Solving the query *p(h)* ⊗ *p*(c)⊗ *p*(c) would also succeed as we can obtain a hamburger for three dollars, a coke and a (refilled) coke for one dollar. The examples presented here have been of a simple nature. They are, however, sufficient for appreciating the attractiveness of the algorithm development process proposed here. We point the reader to [1],[4],[5] for more examples.

CONCLUSION

A proposal for designing algorithms is given. It is based on the view that softwares are dynamic systems simulated on a machine and an algorithm is a constructive definition of a dynamic system. The advantage of our approach is that is simplifies the process of designing and writing algorithms for the problems that require nondeterministic updates. Our ultimate interest is in a procedure for carrying out computations of the kind described above. Hence it is important to realize this linear logic this interpreter in an efficient way, as discussed in [2][4]. In the future, we are also interested in choosing an extension to linear logic, computability logic [7, 8] to express algorithms.

ACKNOWLEDGMENT

This paper was supported by Dong-A University Research Fund in 2009.